\documentclass[twocolumn, published]{aastex701}
\newcommand{\fnDtThreeSigma}{3.50\times10^{2}}

\newcommand{\fnDtFiveSigma}{7.43\times10^{-2}}

\newcommand{\fnPTargetThreeSigma}{1.35\times10^{-3}}

\newcommand{\fnPTargetFiveSigma}{2.87\times10^{-7}}

\newcommand{\fnPMCObs}{0.254}

\newcommand{\fnSigmaMCObs}{0.66}

\newcommand{\fnExpectedExceedThreeSigma}{1.35\times10^{5}}

\newcommand{\fnRelPrecisionThreeSigma}{0.27}

\newcommand{\fnObservedCountsThreeSigma}{134550}

\newcommand{\fnExpectedExceedFiveSigma}{28.7}

\newcommand{\fnRelPrecisionFiveSigma}{18.68}

\newcommand{\fnObservedCountsFiveSigma}{26}

\newcommand{\fnNMCProduction}{10^{8}}

\newcommand{\fnDtClosureBreakdown}{5.14\times10^{3}}

\newcommand{\fnDtThreeSigmaBinSysLo}{3.290\times10^{2}}

\newcommand{\fnDtThreeSigmaBinSysLoRelPct}{-6.0}

\newcommand{\fnDtThreeSigmaBinSysHi}{4.226\times10^{2}}

\newcommand{\fnDtThreeSigmaBinSysHiRelPct}{20.7}

\newcommand{\fnDtFiveSigmaBinSysLo}{6.982\times10^{-2}}

\newcommand{\fnDtFiveSigmaBinSysLoRelPct}{-6.0}

\newcommand{\fnDtFiveSigmaBinSysHi}{8.967\times10^{-2}}

\newcommand{\fnDtFiveSigmaBinSysHiRelPct}{20.7}

\newcommand{\fnNFRBTotal}{1823}
\newcommand{\fnNFRBTotalAfterFlags}{1662}
\newcommand{\fnNFRBNonRepeating}{1423}
\newcommand{\fnNFRBValidCoordinates}{1423}
\newcommand{\fnNFRBLimitsXi}{1332}
\newcommand{\fnNFRBNonRepeatingRaw}{1562}

\newcommand{\fnNFlagExcluded}{145}
\newcommand{\fnNFlagSidelobe}{9}
\newcommand{\fnNFlagCitizen}{13}
\newcommand{\fnNFlagUnion}{161}

\newcommand{\fnNCommissioning}{13}

\newcommand{\fnRateBinDays}{30}
\newcommand{\fnRateFirstEdgeMJD}{58324}
\newcommand{\fnRateLastEdgeMJD}{59194}

\newcommand{\fnFigOneDeltaTDays}{0.9342}

\newcommand{\fnNAlertsNIZero}{40}

\newcommand{\fnXiULOne}{1.4\times10^{11}}

\newcommand{\fnFluencePivotTwo}{5.28\times10^{-4}}

\newcommand{\fnXiULThree}{6.8\times10^{12}}

\newcommand{\fnNeffW}{190}

\newcommand{\fnNeffK}{1136}

\newcommand{\fnNeffWRowC}{12.6}

\newcommand{\fnGPhot}{6.48}

\newcommand{\fnSSamp}{0.094}

\newcommand{\fnNBasebandUsed}{95}
\newcommand{\fnCRatioMedian}{3.35}

\newcommand{\fnNCRatioBelowOne}{4}

\newcommand{\fnJackknifeAlert}{IC190629A}
\newcommand{\fnJackknifeFSharePercent}{52.8}

\newcommand{\fnJackknifeRatio}{2.12}

\newcommand{\fnNPairs}{11}

\newcommand{\fnPairWindowDays}{30}
\newcommand{\fnNAlertsInPairs}{7}
\newcommand{\fnNPairsLargestAlert}{5}
\newcommand{\fnNAlerts}{73}

\newcommand{\fnRatioDtThreeSigma}{231}

\newcommand{\fnNextPairDays}{1.60}

\newcommand{\fnNCatalogue}{4539}
\newcommand{\fnNSourcesCatalogue}{3641}
\newcommand{\fnFracPeriodPct}{40}
\newcommand{\fnNRepeaters}{239}

\newcommand{\fnNNoFluence}{91}

\newcommand{\fnNPairsExpected}{8.7}

\newcommand{\fnNPairsExpectedStd}{2.5}

\newcommand{\fnObsYears}{2.4}

\newcommand{\fnQiMax}{0.4398}

\newcommand{\fnNiIC}{0.0007129}

\newcommand{\fnDtThreeSigmaNoFlag}{3.42\times10^{2}}

\newcommand{\fnDtFiveSigmaNoFlag}{7.25\times10^{-2}}

\newcommand{\fnBasebandFracPct}{7}

\newcommand{\fnPriorRatioFahey}{50}

\newcommand{\fnPriorRatioAlbert}{2083}

\newcommand{\fnPriorRatioAartsenA}{4.0}

\newcommand{\fnPriorRatioAartsenB}{3.8}

\newcommand{\fnPriorRatioAbbasi}{38}

\newcommand{\fnRedshiftReference}{1.2}

\newcommand{\fnRedshiftFractionBelowReferencePct}{90}

\newcommand{\fnKCorrMaxFactorAtReference}{4.8}

\newcommand{\fnPivotEnergyTeV}{100}

\newcommand{\fnPeriodStart}{2018 July 25}

\newcommand{\fnPeriodEnd}{2020 December 31}

\newcommand{\fnMuNinety}{2.30}
\newcommand{\fnNMeanAtThreeSigma}{1.4\times10^{-3}}

\newcommand{\fnNMeanAtFiveSigma}{2.9\times10^{-7}}

\begin{document}

\title[FRB--neutrino associations]{Constraints on the Physical Association between ICECAT1 Neutrinos and Fast Radio Bursts Using the Second CHIME/FRB Catalogue}
\shorttitle{FRB--Neutrino Association Search}
\shortauthors{Masaoka et al.}
\submitjournal{ApJ}


\correspondingauthor{Hiroto Masaoka}
\email{hmasaoka@g.ecc.u-tokyo.ac.jp}

\author[orcid=0009-0003-5746-1510, gname=Hiroto,sname=Masaoka]{Hiroto Masaoka}
\email{hmasaoka@g.ecc.u-tokyo.ac.jp}
\affiliation{Department of Astronomy, School of Science, The University of Tokyo, 7-3-1 Hongo, Bunkyo-ku, Tokyo 113-0033, Japan}
\affiliation{Mizusawa VLBI Observatory, National Astronomical Observatory of Japan, 1-2-2 Mizusawa-Hoshigaoka, Oshu, Iwate 023-0861, Japan}

\author[orcid=0000-0001-7228-1428, gname=Tetsuya,sname=Hashimoto]{Tetsuya Hashimoto}
\email{tetsuya@phys.nchu.edu.tw}
\affiliation{Department of Physics, National Chung Hsing University, No. 145, Xingda Rd., South Dist., Taichung, 40227, Taiwan}

\author[orcid=0000-0002-1688-8708, gname=Shotaro,sname=Yamasaki]{Shotaro Yamasaki}
\email{shotaro.s.yamasaki@gmail.com}
\affiliation{Department of Physics, National Chung Hsing University, No. 145, Xingda Rd., South Dist., Taichung, 40227, Taiwan}

\author[orcid=0000-0002-9255-4742, gname=Yuhei,sname=Iwata]{Yuhei Iwata}
\email{yuhei.iwata@nao.ac.jp}
\affiliation{Mizusawa VLBI Observatory, National Astronomical Observatory of Japan, 1-2-2 Mizusawa-Hoshigaoka, Oshu, Iwate 023-0861, Japan}
\affiliation{Astronomical Science Program, Graduate Institute for Advanced Studies, SOKENDAI, 2-21-1 Osawa, Mitaka, Tokyo 181-8588, Japan}

\author[orcid=0000-0001-6010-714X, gname=Tomoki,sname=Wada]{Tomoki Wada}
\email{tomoki.wada@astr.tohoku.ac.jp}
\affiliation{Department of Physics, National Chung Hsing University, No. 145, Xingda Rd., South Dist., Taichung, 40227, Taiwan}
\affiliation{Frontier Research Institute for Interdisciplinary Sciences, Tohoku University, Sendai, Miyagi 980-8578, Japan}
\affiliation{Astronomical Institute, Graduate School of Science, Tohoku University, Sendai, Miyagi 980-8578, Japan}

\author[orcid=0000-0002-6821-8669, gname=Tomotsugu,sname=Goto]{Tomotsugu Goto}
\email{itlhnis@gmail.com}
\affiliation{Institute of Astronomy, National Tsing Hua University, No. 101, Section 2, Kuang-Fu Road, Hsinchu City 30013, Taiwan}

\author[orcid=0000-0003-0581-5973, gname=Shintaro,sname=Yoshiura]{Shintaro Yoshiura}
\email{shintaro.yoshiura@nao.ac.jp}
\affiliation{Mizusawa VLBI Observatory, National Astronomical Observatory of Japan, 2-21-1 Osawa, Mitaka, Tokyo 181-8588, Japan}

\author[orcid=0009-0001-7459-1670, gname=Kazuaki,sname=Hashiyama]{Kazuaki Hashiyama}
\email{kazuaki.hashiyama@nao.ac.jp}
\affiliation{Mizusawa VLBI Observatory, National Astronomical Observatory of Japan, 1-2-2 Mizusawa-Hoshigaoka, Oshu, Iwate 023-0861, Japan}

\author[orcid=0000-0003-4058-9000, gname=Mareki,sname=Honma]{Mareki Honma}
\email{mareki.honma@nao.ac.jp}
\affiliation{Department of Astronomy, School of Science, The University of Tokyo, 7-3-1 Hongo, Bunkyo-ku, Tokyo 113-0033, Japan}
\affiliation{Mizusawa VLBI Observatory, National Astronomical Observatory of Japan, 1-2-2 Mizusawa-Hoshigaoka, Oshu, Iwate 023-0861, Japan}
\affiliation{Mizusawa VLBI Observatory, National Astronomical Observatory of Japan, 2-21-1 Osawa, Mitaka, Tokyo 181-8588, Japan}

\author[orcid=0000-0001-9399-5331, gname=Takuya,sname=Akahori]{Takuya Akahori}
\email{takuya.akahori@nao.ac.jp}
\affiliation{Mizusawa VLBI Observatory, National Astronomical Observatory of Japan, 2-21-1 Osawa, Mitaka, Tokyo 181-8588, Japan}

\author[orcid=0000-0003-2955-1239, gname=Kohei,sname=Kurahara]{Kohei Kurahara}
\email{kurahara.kohei.i7@f.mail.nagoya-u.ac.jp}
\affiliation{Institute for the Origin of Particles and the Universe (KMI), Nagoya University, Furo-cho, Chikusa-ku, Nagoya, Aichi 464-8601, Japan}


\begin{abstract}

We present a search for neutrino counterparts to fast radio bursts (FRBs) using temporal and spatial cross-matching between the $\fnNFRBNonRepeating$ non-repeating FRBs of the Second CHIME/FRB catalogue and the $\fnNAlerts$ alerts of the IceCube high-energy alert-track catalogue ICECAT1 from \fnPeriodStart{} to \fnPeriodEnd{}. Because current FRB--neutrino models do not agree on the emission order, we adopt a two-sided temporal hypothesis and define the significance criterion from the expected number of chance coincidences before the search. We find no statistically significant association: no pair lies within the $3\sigma$ and $5\sigma$ coincidence windows of $\fnDtThreeSigma$~s and $\fnDtFiveSigma$~s, and the pair closest in time, FRB\,20190630C--IC\,190629A, is consistent with a chance coincidence ($p=\fnPMCObs$, $\fnSigmaMCObs\,\sigma$). Using a population-level stacking analysis, we derive 90\% upper limits on the neutrino-to-radio luminosity ratio of FRBs, $\xi \lesssim \fnXiULOne$ for a neutrino spectral index $\gamma=1.0$ to $\fnXiULThree$ for $\gamma=3.0$, assuming neutrino emission within the burst duration. For a neutrino number spectrum $dN/d\epsilon \propto \epsilon^{-2}$, the corresponding limit on the neutrino fluence per burst is about a factor of 4 below those of previous stacking searches. 
For $\gamma=2$, the limits on the isotropic neutrino energy, $10^{49}$--$10^{52}$~erg, lie more than five orders of magnitude above the predictions of magnetar-driven shock models, so the present data do not constrain these models. Nevertheless, our blind significance criterion and population-stacking limits provide a baseline for future multi-messenger searches.
    
\end{abstract}


\section{Introduction}
\label{section1}

Fast Radio Bursts (FRBs) are extraordinarily powerful radio transients characterized by a brightness temperature of $\sim 10^{36} \,\mathrm{K}$, millisecond durations, and large dispersion measures (DM), indicating their origin from cosmological distances \citep[e.g.,][]{Lorimer_2007}. Despite significant observations and theoretical progress \citep[e.g.][]{Petroff_2022}, the progenitors and mechanisms of FRBs remain elusive. To distinguish between various FRB models proposed so far \citep[e.g.,][]{Platts_2019}, acquiring additional insights through multi-wavelength and multi-messenger observations is essential. For example, the detection of FRB\,20200428D \citep{CHIME_2020, Bochenek_2020} and its association with hard X-ray bursts \citep{Mereghetti_2020, Li_2021, Ridnaia_2021, Tavani_2021, Li_2022} strongly supports magnetars as plausible FRB progenitors.

Despite extensive follow-up efforts \citep[e.g.][]{Yamasaki_2016,Nicastro_2021, Zhang_2024}, identifying high-energy counterparts to extragalactic FRBs remains challenging. A 15--50~keV GRB-like signal lasting for 300~s was tentatively linked to FRB\,20131104 at $3.2\sigma$ significance \citep{DeLaunay_2016}. A potential association between the binary neutron star merger GW\,190425 \citep{Abbott_2020} and the non-repeating FRB\,20190425A was reported with a chance probability of 0.0052 ($2.8\sigma$) \citep{Moroianu_2022}. Additionally, recent theoretical work has proposed high-energy neutrino emission in the TeV--PeV range associated with FRBs via photo-hadronic interactions \citep{Metzger_2020, Qu_2022, Shimoda_2024}, with different delay structures and possible emission orderings (FRB-before-neutrino or FRB-after-neutrino). However, despite multiple searches using ANTARES and IceCube data, no statistically significant neutrino counterparts to FRBs have been identified. In small-sample studies (4--29 FRBs), temporal and spatial coincidence criteria were applied simultaneously but with limited event statistics \citep{Fahey_2017, Albert_2018a, Aartsen_2018c, Aartsen_2020a, Abbasi_2023c}, whereas the larger-sample studies tested either spatial coincidence or the time-integrated flux \citep{Desai_2023, Luo_2024}. This methodological split has made it difficult to establish a robust physical FRB--neutrino connection.

More generally, the identification of astrophysical neutrino sources is challenged by substantial atmospheric background noise and limited directional accuracy, which weakens source-association significance between neutrino detections and their potential sources \citep[e.g.][]{Troitsky_2021a, Sharma_2024a}. Despite the hundreds of high-energy neutrino detections by the IceCube neutrino observatory in the TeV--PeV energy range \citep{Abbasi_2023b}, only a small number of promising candidate sources have been identified. In 2018, the IceCube collaboration reported an association between a high-energy neutrino event, IC\,170922A, and a gamma-ray flare from blazar TXS\,0506+056, with a significance of $3~\sigma$ \citep{Aartsen_2018b}. They also reported another association between the blazar TXS\,0506+056 and a neutrino flare in the same direction between 2014 September and 2015 March, during which TXS\,0506+056 was not at a gamma-ray flaring state \citep{Aartsen_2018a}. This association significance was $2.4~\sigma$ \citep{Abbasi_2021a, Abbasi_2023a}. In 2022, the IceCube collaboration found evidence of neutrino emission from a type 2 Seyfert galaxy, NGC\,1068, with a significance of $4.2~\sigma$ \citep{Abbasi_2022a}. Both findings have been associated with the detection of multiple neutrinos across a range of energies. In contrast, the significance of other candidate sources, typically identified based only on a single neutrino detection, is weaker \citep[][Sec.~VI]{Bustamante_2024a}. These previous works illustrate that modest-significance ($\sim$2--4~$\sigma$) associations can provide useful exploratory clues in multi-messenger astronomy, but establishing a physical source class requires additional events and independent confirmation.

In this paper, aiming to search for neutrino counterparts to FRBs, we test the hypothesis that FRBs are accompanied by high-energy neutrino emission close in both time and sky position, using neutrino alerts from ICECAT1 \citep{Abbasi_2023b} and FRBs from the Second CHIME/FRB Catalogue \citep{Abbott_2026}. 
Unlike the previous larger-sample studies \citep{Desai_2023, Luo_2024}, which tested spatial coincidence or the time-integrated flux alone, our test requires the neutrino to arrive close in time to the FRB. Three elements characterize our analysis. First, we use a large FRB sample from a single instrument, covering the full period in which the FRB and neutrino data overlap. Second, we define the significance criterion from the expected number of chance coincidences independently of the pairs found in the search (a blind analysis). Third, we stack the bursts to constrain the neutrino-to-radio luminosity ratio of the FRB population.

This paper is structured as follows. In Section~\ref{sec:data}, we describe the FRB sample and the neutrino alerts. In Section~\ref{sec:blind}, we define the significance criterion independently of the search. In Section~\ref{sec:results}, we present the coincidence search, and in Section~\ref{sec:limits}, we derive the upper limits on the neutrino-to-radio luminosity ratio and on the neutrino fluence per burst. In Section~\ref{sec:Discussion}, we examine the dependence on the sample definition, compare the results with previous works, discuss the implications for neutrino emission models, and describe prospects. We summarize our conclusions in Section~\ref{sec:summary}.

\section{Data and event selection}
\label{sec:data}
\label{subsec:period}

We search for coincidences between FRBs from the second CHIME/FRB catalogue \citep{Abbott_2026} and neutrino alerts from ICECAT1 \citep{Abbasi_2023b} over the period in which the two overlap, from \fnPeriodStart{}, the start of the catalogue, to \fnPeriodEnd{}, the end of ICECAT1 (MJD 58324--59214, $\fnObsYears$~yr).
We refer to this as the analysis period.

\subsection{FRB sample}
\label{subsec:frbsample}

The catalogue contains $\fnNCatalogue$ bursts from $\fnNSourcesCatalogue$ sources up to 2023 September 15, of which $\fnNFRBTotal$ ($\fnFracPeriodPct$\%) fall in the analysis period (Table~\ref{tab:frbsample}).
A burst with several sub-bursts is treated as a single event, and its arrival time is that of the first sub-burst (column \texttt{mjd\_inf}: topocentric, referenced to infinite frequency, in MJD UTC).
As recommended for statistical analyses by \citet{Abbott_2026}, we exclude the $\fnNFlagUnion$ bursts that carry at least one of the flags \texttt{excluded\_flag} ($\fnNFlagExcluded$ bursts), \texttt{sidelobe\_flag} ($\fnNFlagSidelobe$) and \texttt{citizen\_science\_flag} ($\fnNFlagCitizen$); some bursts carry two flags.
The $\fnNCommissioning$ bursts detected during commissioning are removed by \texttt{excluded\_flag}.
No S/N threshold is applied.

Of the remaining $\fnNFRBTotalAfterFlags$ bursts, we use the $\fnNFRBNonRepeating$ apparently non-repeating FRBs (no \texttt{repeater\_name}), because bursts from one repeater are not independent trials.
All of them have a sky position and enter the coincidence search.
The stacking limit also requires a fluence, which $\fnNNoFluence$ of them lack; since the FRBs without a fluence need not be a random subset, the limit applies to the $\fnNFRBLimitsXi$ FRBs that have one.
The effects of the flag exclusion and of including the repeaters are examined in Section~\ref{sec:robust}.
The catalogue positions come from intensity data; the first CHIME/FRB baseband catalogue \citep{Amiri_2024a}, which gives more precise localizations and fluences for a subset of the bursts, is not needed for the search, since the alert regions are much larger, and is used only to check the fluence calibration (Section~\ref{sec:fluence-cal}).

\begin{deluxetable*}{lrll}
\tablecaption{FRB samples used in this work\label{tab:frbsample}}
\tablehead{
\colhead{Sample} & \colhead{$N$} & \colhead{Condition} & \colhead{Use}
}
\startdata
Catalogue & \fnNCatalogue & all bursts up to 2023 September 15 & --- \\
Analysis period & \fnNFRBTotal & MJD 58324--59214 & --- \\
\quad Non-repeating FRBs, no flag exclusion & \fnNFRBNonRepeatingRaw & no \texttt{repeater\_name} & comparison without flag exclusion \\
\quad Flag-excluded & \fnNFRBTotalAfterFlags & three quality flags excluded\tablenotemark{a} & comparison with repeaters included \\
\qquad Non-repeating FRBs & \fnNFRBNonRepeating & no \texttt{repeater\_name} & coincidence search (baseline) \\
\qquad Repeaters & \fnNRepeaters & with \texttt{repeater\_name} & --- \\
\qquad With fluence & \fnNFRBLimitsXi & non-repeating, with fluence & stacking limit \\
\enddata
\tablenotetext{a}{\texttt{excluded\_flag}, \texttt{sidelobe\_flag} and \texttt{citizen\_science\_flag}}
\tablecomments{Indented rows are subsets of the row above their indentation level; the sample with fluence is a subset of the non-repeating FRBs of the flag-excluded sample. The comparison samples are used in Section~\ref{sec:robust}.}
\end{deluxetable*}

\subsection{Neutrino alerts}
\label{subsec:nudata}
\label{subsec:selection}

We test the hypothesis that FRBs are accompanied by high-energy neutrino emission, which would place neutrino alerts close to FRBs in both time and sky position.
ICECAT1 \citep{Abbasi_2023b} suits this test for three reasons.
First, it overlaps the CHIME/FRB catalogue in time, whereas the public 10-yr point-source data \citep[IceTracks-DR1;][]{Abbasi_2021a} end in 2018 July, when CHIME/FRB started operating. 
Second, it gives the effective areas of the Gold and Bronze alert selections, from which the expected number of neutrinos from each FRB can be calculated (Section~\ref{sec:limits}).
Third, it gives a localization region for each alert, whose size alone determines the expected number of chance coincidences, without using the arrival times of the neutrinos (Section~\ref{sec:blind}).

ICECAT1 contains $\fnNAlerts$ alerts in the analysis period, from both the Gold and Bronze streams, each with a best-fit position and asymmetric 90\% containment uncertainties.
ICECAT1 gives a signalness for each alert.
A cut or weight on signalness could raise the sensitivity to astrophysical neutrinos, but we use all alerts with equal weight, for three reasons.
First, the published effective areas refer to the full Gold and Bronze selections, so the expected number of neutrinos cannot be calculated for a subset selected on signalness.
Second, a signalness threshold would be a free parameter, and choosing it after examining the data would introduce an additional trial.
Third, signalness is a posterior probability that depends on an assumed diffuse astrophysical flux, and weighting by it would make the result conditional on that assumption.


\section{Blind Analysis and the Significance Criterion}
\label{sec:blind}

This section defines the significance criterion from the probability that unrelated FRBs and neutrino alerts coincide by chance.
The probability is built only from properties of the two catalogues: the detection rate of CHIME/FRB in a given declination range and $\fnRateBinDays$-day interval, and, for each neutrino alert, the right-ascension width of its localization region as a fraction of $360^\circ$.
It does not use the arrival-time difference of any FRB--neutrino pair, so the significance criterion and the configuration of the stacking analysis (Section~\ref{sec:limits}) are defined independently of the pairs found in the association search of Section~\ref{sec:results}; the analysis is \emph{blind by construction}.

\subsection{Null Hypothesis: the Scrambling Ensemble}
\label{sec:null}

The null hypothesis is that FRBs and neutrino alerts are unrelated.
We simulate it by scrambling the neutrino alerts and leaving the FRB catalogue unchanged.
In each realization, every neutrino alert keeps its declination and 90\% localization region, but its right ascension is drawn uniformly from $[0^\circ,360^\circ)$ and its arrival time uniformly within the $\fnRateBinDays$-day interval in which it was detected.
Uniform right-ascension scrambling is justified by IceCube's approximately uniform time-integrated exposure in right ascension, owing to Earth's rotation at the South Pole; the declination dependence of the effective area is preserved because each alert keeps its declination.

\begin{figure}
\centering
\includegraphics[width=\columnwidth]{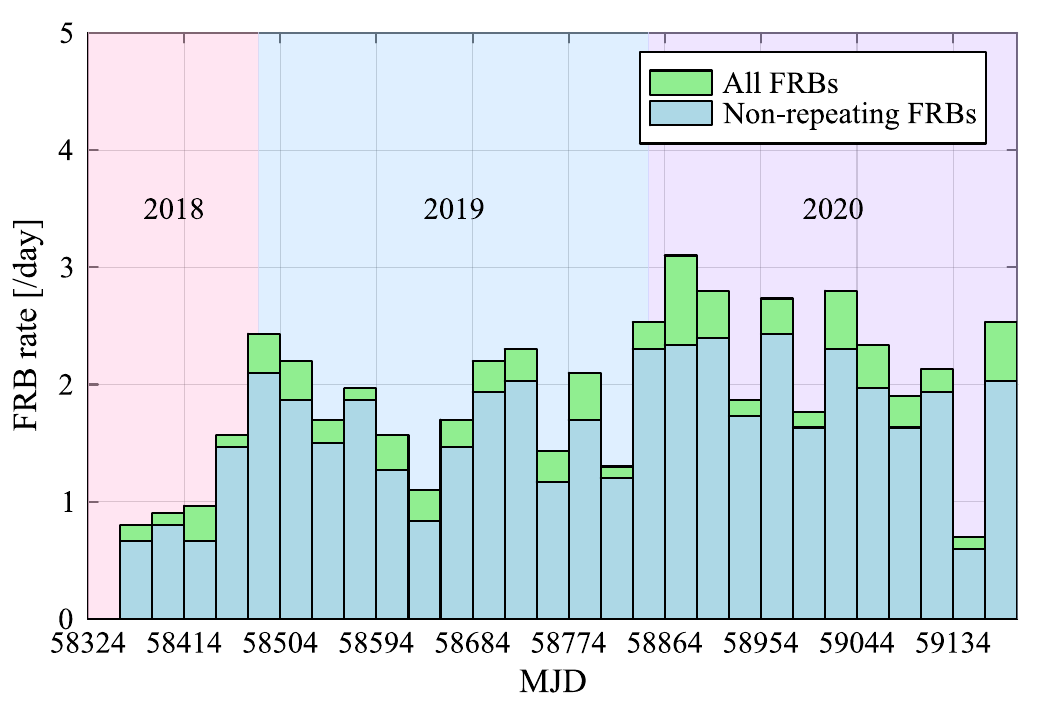}
\caption{Average daily rate of FRBs in the flag-excluded sample of the Second CHIME/FRB Catalogue, binned in $\fnRateBinDays$-day intervals over MJD $\fnRateFirstEdgeMJD$--$\fnRateLastEdgeMJD$ (2018 July 25 to 2020 December 11).
Light blue: non-repeating FRBs; light green: all FRBs, including repeaters, with one entry per TNS name.
Shading marks calendar years. The histogram shows the all-sky rate, whereas $R_i$ includes only FRBs within the declination range $D_i$.}
\label{fig:CHIME_FRB_Rate}
\end{figure}

We scramble the neutrino alerts rather than the FRBs to keep the FRB catalogue as observed, including the transit design of CHIME, which couples the right ascension of a burst to its detection time.
The arrival times are randomized only within $\fnRateBinDays$-day intervals and not over the full $\fnObsYears$~yr, so that the realizations follow the long-term variation of the CHIME/FRB detection rate (Figure~\ref{fig:CHIME_FRB_Rate}) while each interval contains enough FRBs to estimate it.

\subsection{Expected Number of Chance Coincidences}
\label{sec:expected}

Consider a neutrino alert $\nu_i$ and FRBs detected within $\pm\Delta T$ of its arrival time.
We adopt a two-sided time window because the relative emission order remains unconstrained in proposed FRB--neutrino emission models \citep[e.g.,][]{Metzger_2020, Qu_2022}.
We approximate the 90\% localization region of $\nu_i$ by the rectangle in equatorial coordinates given by the asymmetric uncertainties reported in ICECAT1, which bound the 90\% contour \citep{Abbasi_2023b}.
The rectangle spans a declination range $D_i$ and a right-ascension width $\Delta\alpha_i$ (a coordinate difference, not an angle on the sky).
It encloses the region, so the expected number of chance coincidences is slightly overestimated; the coincidence search (Section~\ref{sec:results}) uses the same rectangle.

The expected number of FRBs inside the rectangle and within $\pm\Delta T$ of the neutrino arrival time is
\begin{equation}
  \langle n_i\rangle(\Delta T) = 2\,\Delta T\,R_i\,q_i,
  \label{eq:ni}
\end{equation}
where $R_i$ is the mean detection rate of non-repeating FRBs at any right ascension within $D_i$ during the $\fnRateBinDays$-day interval containing the neutrino detection, and $q_i\equiv\Delta\alpha_i/360^\circ$ is the probability that an FRB falls inside the rectangle in right ascension, because the right ascension of a scrambled alert is uniform (Figure~\ref{fig:qi_DEC}).
For alerts whose right-ascension uncertainty exceeds $360^\circ$, we set $q_i=1$.

\subsection{Probability of at Least One Chance Coincidence}
\label{sec:prob}

Summing Equation~(\ref{eq:ni}) over the $\fnNAlerts$ neutrino alerts, the total expected number of chance coincidences is given by
\begin{equation}
  \langle N\rangle(\Delta T) = \sum_{i=1}^{\fnNAlerts}\langle n_i\rangle(\Delta T) = 2\,\Delta T\sum_{i=1}^{\fnNAlerts}R_iq_i .
  \label{eq:N}
\end{equation}
If non-repeating FRBs follow a Poisson process, with which they are consistent \citep{Ameri_2020a, Rajwade_2020a}, and the coincidence windows for different neutrino alerts do not overlap, the total count follows a Poisson distribution with mean $\langle N\rangle$, and the probability that the background produces at least one chance coincidence within $\pm\Delta T$ is given by
\begin{equation}
  p(\Delta T) = 1-\exp\left[-\langle N\rangle(\Delta T)\right].
  \label{eq:p}
\end{equation}
Because the sum runs over all neutrino alerts, $p$ is a global probability that needs no further correction for their multiplicity.
Deviations from a Poisson distribution introduce a correction of relative order $\langle N\rangle$ to Equation~(\ref{eq:p}), which we quantify in Section~\ref{sec:blind-sens}.

\citet{Moroianu_2022} estimate the chance probability as the product $P_TP_S$ of temporal and spatial factors, an approximation valid when the expected number of coincidences is much smaller than unity.
We instead use the Poisson expression of Equation~(\ref{eq:p}), which accounts for the saturation of $p$ at large $\langle N\rangle$ and therefore covers a wide range of $\Delta T$.

\subsection{Blind Coincidence Windows and Closure Test}
\label{sec:blind-sens}

\begin{figure}
\centering
\includegraphics[width=\columnwidth]{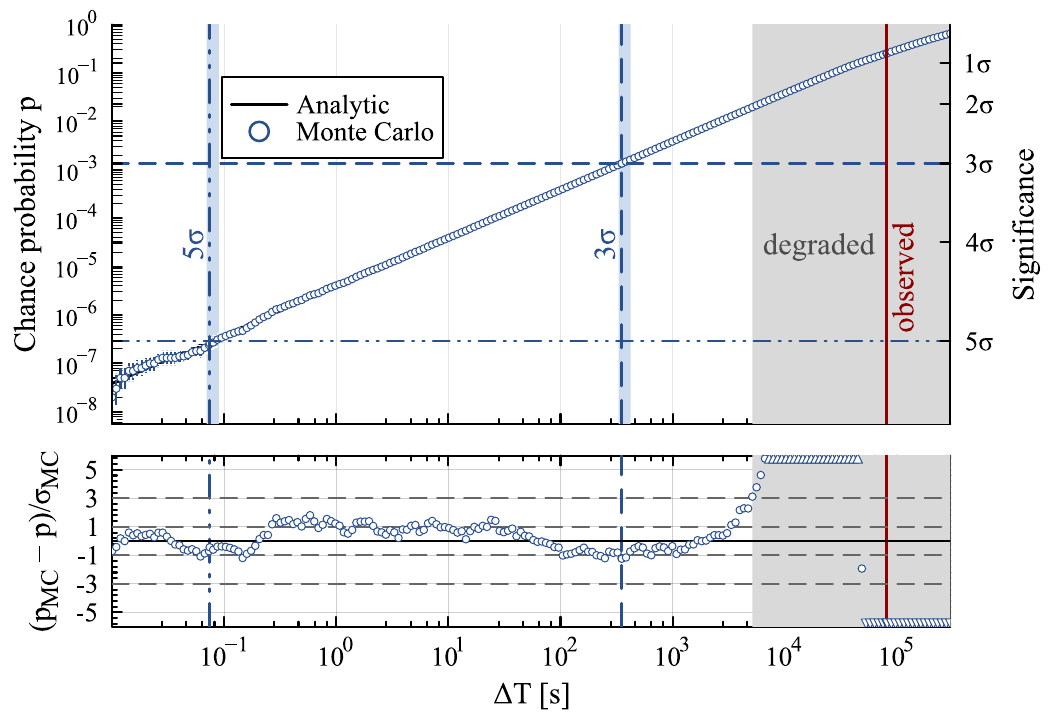}
\caption{Chance-coincidence probability versus coincidence window $\Delta T$.
\emph{Upper panel:} the black solid curve is Equation~(\ref{eq:p}), dotted beyond the closure breakdown (shaded). Open circles show the fraction of the $\fnNMCProduction$ Monte Carlo realizations with $\mathrm{TS}\le\Delta T$, with $68.27\%$ Clopper--Pearson intervals \citep{Clopper_Pearson_1934}; the right axis gives the equivalent one-sided Gaussian significance. Dashed and dash-dot-dot lines mark the $3\sigma$ and $5\sigma$ levels and windows, and the bands show the range of the windows for rate-bin widths of 10--60 days. The red vertical line marks the closest FRB--neutrino pair of Section~\ref{sec:results} ($\Delta T=\fnFigOneDeltaTDays$~d).
\emph{Lower panel:} the pull $(p_{\rm MC}-p)/\sigma_{\rm MC}$, where $\sigma_{\rm MC}$ is the half-width of the Clopper--Pearson interval on the side facing Equation~(\ref{eq:p}); lines are at $0$, $\pm1$ and $\pm3$, and triangles mark pulls beyond the axis range. The points are cumulative counts and therefore strongly correlated.}
\label{fig:closure}
\end{figure}

We define the significance criterion at the one-sided Gaussian $3\sigma$ ($p=\fnPTargetThreeSigma$) and $5\sigma$ ($p=\fnPTargetFiveSigma$) levels.
The $3\sigma$ and $5\sigma$ coincidence windows are the values of $\Delta T$ at which Equation~(\ref{eq:p}) reaches these probabilities; inverting it gives $\Delta T_{3\sigma}=\fnDtThreeSigma$~s and $\Delta T_{5\sigma}=\fnDtFiveSigma$~s (Figure~\ref{fig:closure}).
The criterion is met if an FRB--neutrino pair lies within one of these windows.
We validate these windows against the Monte Carlo realizations of the scrambling ensemble, and Table~\ref{tab:thresholds} summarizes the results.

For each realization, we define the test statistic as
\begin{equation}
  \mathrm{TS}\equiv\min\left\{|t_j-t_i| : i=1,\ldots,\fnNAlerts,\ j \in \mbox{region of alert }i\right\},
  \label{eq:TS}
\end{equation}
where $t_i$ and $t_j$ are the arrival time of neutrino alert $\nu_i$ and the detection time of FRB $j$ within its localization region, respectively.
A coincidence occurs within $\pm\Delta T$ if and only if $\mathrm{TS}\le\Delta T$, making the cumulative distribution of $\mathrm{TS}$ the empirical counterpart of Equation~(\ref{eq:p}).
Since $\mathrm{TS}$ is already the minimum over all neutrino alerts and FRBs, the choice of $\Delta T$ introduces no additional trials factor.
The probability that the null hypothesis produces a $\mathrm{TS}$ at most as large as the observed one, $p(\mathrm{TS_{obs}})$, is therefore the $p$-value of the test, and it includes the multiplicity of the alerts and the FRBs.
The significance criterion is equivalent to $p(\mathrm{TS_{obs}})<\fnPTargetThreeSigma$ ($3\sigma$) or $<\fnPTargetFiveSigma$ ($5\sigma$).

Figure~\ref{fig:closure} compares Equation~(\ref{eq:p}) with the cumulative distribution of $\mathrm{TS}$ from $\fnNMCProduction$ scrambled realizations.
The two agree within $3\sigma_{\rm MC}$, where $\sigma_{\rm MC}$ is the Monte Carlo uncertainty, below the closure breakdown, which we define as the smallest $\Delta T$ at which they differ by more than $3\sigma_{\rm MC}$.
The closure breakdown is $\Delta T\simeq\fnDtClosureBreakdown$~s, well beyond both windows.
At larger $\Delta T$, the Monte Carlo results deviate from the prediction; this does not affect the windows, and we use the Monte Carlo results beyond the closure breakdown.

The windows depend on the rate-bin width, which sets how finely $R_i$ follows the time variation of the CHIME/FRB detection rate.
Varying it from 10 to 60 days changes the windows by $\fnDtThreeSigmaBinSysLoRelPct\%$ to $+\fnDtThreeSigmaBinSysHiRelPct\%$, which we take as their systematic uncertainty; it is shown as the bands in Figure~\ref{fig:closure} and listed in Table~\ref{tab:thresholds}.

The other sources of uncertainty in the windows are smaller than this, or comparable at $5\sigma$.
The Monte Carlo reproduces Equation~(\ref{eq:p}) with a fractional precision of $\fnRelPrecisionThreeSigma\%$ at $3\sigma$ and $\fnRelPrecisionFiveSigma\%$ at $5\sigma$, which limits the agreement at the windows.
The departure of Equation~(\ref{eq:p}) from the exact probability is of relative order $\langle N\rangle$ (Section~\ref{sec:prob}), which is $\fnNMeanAtThreeSigma$ at $3\sigma$ and $\fnNMeanAtFiveSigma$ at $5\sigma$.
We therefore adopt the windows derived from Equation~(\ref{eq:p}) rather than those from the Monte Carlo.
At $5\sigma$, the Monte Carlo cannot test Equation~(\ref{eq:p}) directly; the agreement at larger $\Delta T$, where the departure is larger, supports its use there.

\begin{figure}
\centering
\includegraphics[width=\columnwidth]{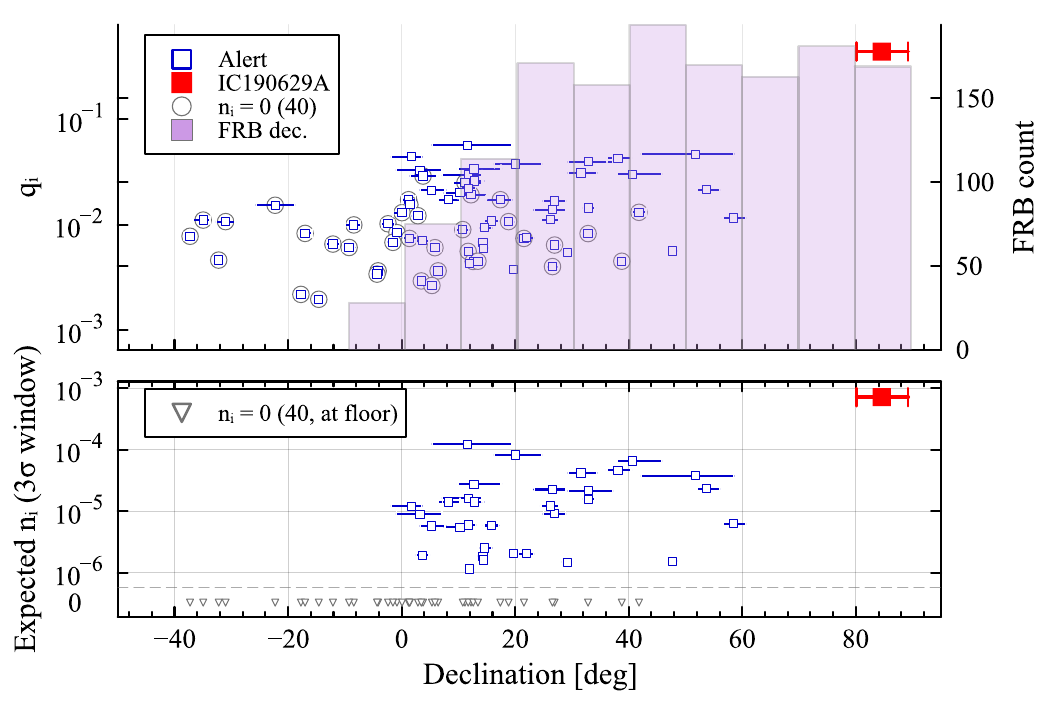}
\caption{Upper panel: right-ascension fraction $q_i$ of the $\fnNAlerts$ neutrino alerts (open squares; left axis) versus declination, with the declination uncertainties as horizontal error bars. The purple histogram (right axis) shows the declination distribution of the $\fnNFRBNonRepeating$ non-repeating FRBs. Lower panel: expected number of chance coincidences $\langle n_i\rangle$ per alert at the $3\sigma$ window ($\Delta T_{3\sigma}=\fnDtThreeSigma$~s). Grey symbols mark the $\fnNAlertsNIZero$ alerts with $R_i=0$, for which $\langle n_i\rangle=0$. IC190629A (red) has the largest $q_i=\fnQiMax$ and the largest $\langle n_i\rangle=\fnNiIC$.}
\label{fig:qi_DEC}
\end{figure}

The windows are also sensitive to individual alerts.
Because the windows are common to all alerts, an alert with a large $R_iq_i$ takes a large share of $\langle N\rangle$ and narrows the windows of all alerts.
\fnJackknifeAlert\ (Figure~\ref{fig:qi_DEC}) alone contributes $\fnJackknifeFSharePercent\%$ of $\langle N\rangle$; without it, the windows would be wider by a factor of $\fnJackknifeRatio$.
This lowers the sensitivity of the criterion, but not its validity, which rests on Equation~(\ref{eq:p}).

\begin{deluxetable}{lcc}
\tabletypesize{\footnotesize}
\tablecaption{The $3\sigma$ and $5\sigma$ coincidence windows from Equation~(\ref{eq:p}) and their uncertainties\label{tab:thresholds}}
\tablehead{
\colhead{Quantity} & \colhead{$3\sigma$} & \colhead{$5\sigma$}
}
\startdata
$\Delta T$, baseline (s) & $\fnDtThreeSigma$ & $\fnDtFiveSigma$ \\
$\Delta T$, no flag excl.\ (s) & $\fnDtThreeSigmaNoFlag$ & $\fnDtFiveSigmaNoFlag$ \\
MC count, expected\tablenotemark{a} & $\fnExpectedExceedThreeSigma$ & \fnExpectedExceedFiveSigma \\
MC count, observed & \fnObservedCountsThreeSigma & \fnObservedCountsFiveSigma \\
\cutinhead{Relative uncertainty of $\Delta T$}
Correction to Eq.~(\ref{eq:p})\tablenotemark{b} & $\fnNMeanAtThreeSigma$ & $\fnNMeanAtFiveSigma$ \\
MC precision (\%) & $\fnRelPrecisionThreeSigma$ & $\fnRelPrecisionFiveSigma$ \\
Rate-bin width (\%)\tablenotemark{c} & $[\fnDtThreeSigmaBinSysLoRelPct,+\fnDtThreeSigmaBinSysHiRelPct]$ & $[\fnDtFiveSigmaBinSysLoRelPct,+\fnDtFiveSigmaBinSysHiRelPct]$ \\
\enddata
\tablenotetext{a}{Number of Monte Carlo realizations with $\mathrm{TS}\le\Delta T$ among $\fnNMCProduction$; the expected number is $\fnNMCProduction p$.}
\tablenotetext{b}{Relative order, estimated by $\langle N\rangle$ at the window.}
\tablenotetext{c}{Windows for 10-, 30-, and 60-day bins relative to the baseline (\fnRateBinDays\ days): $[\fnDtThreeSigmaBinSysLo,\fnDtThreeSigmaBinSysHi]$~s at $3\sigma$ and $[\fnDtFiveSigmaBinSysLo,\fnDtFiveSigmaBinSysHi]$~s at $5\sigma$.}
\end{deluxetable}

\section{Unblinding: the Coincidence Search}
\label{sec:results}
\label{sec:unblind}

This section unblinds the analysis by comparing the arrival times of the FRBs and the neutrino alerts, which the statistic of Section~\ref{sec:blind} does not use.
We test whether any FRB--neutrino pair falls within the $3\sigma$ and $5\sigma$ windows $\Delta T_{3\sigma}$ and $\Delta T_{5\sigma}$ (Section~\ref{sec:blind-sens}) and evaluate the chance probability of the closest pair.

\subsection{Candidate Pairs}
\label{sec:crossmatch}

Table~\ref{table:FRB-Neutrino_Pairs} lists the FRB--neutrino pairs in which the FRB lies inside the 90\% localization region of the alert (the rectangle of Section~\ref{sec:expected}) and $|\Delta T|<\fnPairWindowDays$~days, where $\Delta T\equiv T_{\rm FRB}-T_\nu$ is positive for an FRB detected after the alert.
Among the $\fnNFRBNonRepeating$ non-repeating FRBs, we find $\fnNPairs$ pairs with $\fnNAlertsInPairs$ alerts, including $\fnNPairsLargestAlert$ with IC\,190629A.
This is consistent with the $\fnNPairsExpected\pm\fnNPairsExpectedStd$ pairs expected from the scrambled realizations of Section~\ref{sec:null}.
The $\fnPairWindowDays$-day window is for presentation only: it is much longer than the delays expected from emission models ($\lesssim1000$~s; \citealt{Metzger_2020}) and plays no role in the significance criterion.

\begin{deluxetable*}{cccccccc}
\tablecaption{FRB--neutrino pairs with $|\Delta T|<30$ days}
\label{table:FRB-Neutrino_Pairs}
\tablehead{
\colhead{CHIME FRB} &
\colhead{MJD} &
\colhead{RA} &
\colhead{Dec} &
\colhead{DM} &
\colhead{Fluence} &
\colhead{Coincident Neutrino} &
\colhead{Time Gap} \\
\colhead{} &
\colhead{} &
\colhead{(deg)} &
\colhead{(deg)} &
\colhead{(pc\,cm$^{-3}$)} &
\colhead{(Jy ms)} &
\colhead{} &
\colhead{(day)}
}
\startdata
FRB\,20190409A & 58582.006 & $79.62 \pm 0.20$ & $6.72 \pm 0.27$ & 1791.87 & 8.52 & IC\,190317A & 22.175\\
FRB\,20190622A & 58656.407 & $297.77 \pm 0.23$ & $85.81 \pm 0.29$ & 1122.84 & 1.28 & IC\,190629A & -7.4017\\
FRB\,20190625C & 58659.775 & $73.29 \pm 0.21$ & $11.10 \pm 0.05$ & 442.17 & 4.41 & IC\,190712A & -16.277\\
FRB\,20190630C & 58664.743 & $67.91 \pm 0.21$ & $80.96 \pm 0.09$ & 1660.25 & 1.85 & IC\,190629A & 0.93421\\
FRB\,20190701B & 58665.410 & $302.51 \pm 0.22$ & $80.17 \pm 0.24$ & 749.15 & 2.18 & IC\,190629A & 1.6015\\
FRB\,20190702C & 58666.281 & $284.58 \pm 0.27$ & $88.75 \pm 0.30$ & 1560.94 & 3.87 & IC\,190629A & 2.4723\\
FRB\,20190713B & 58677.967 & $159.85 \pm 0.21$ & $26.48 \pm 0.07$ & 416.06 & 1.09 & IC\,190704A & 9.1831\\
FRB\,20190722B & 58686.703 & $41.09 \pm 0.18$ & $88.97 \pm 0.13$ & 508.43 & 6.84 & IC\,190629A & 22.894\\
FRB\,20200623B & 59023.295 & $257.34 \pm 0.24$ & $26.24 \pm 0.26$ & 259.6 & --- & IC\,200530A & 23.965\\
FRB\,20200813B & 59074.588 & $54.56 \pm 0.21$ & $36.34 \pm 0.21$ & 221.06 & 1.20 & IC\,200911A & -29.010\\
FRB\,20201010A & 59132.793 & $185.62 \pm 0.22$ & $33.73 \pm 0.23$ & 1333.88 & 12.88 & IC\,200926B & 13.852\\
\enddata
\tablecomments{We list FRB name, arrival epoch (MJD UTC, topocentric, referenced to infinite frequency), CHIME right ascension and declination with 1$\sigma$ uncertainties, dispersion measure (\texttt{dm\_fitb}), fluence, associated neutrino event, and time gap. Positive time gaps indicate FRBs detected after the corresponding neutrino alert; negative values indicate FRBs detected before the alert. The pair FRB\,20190609A--IC\,190629A (time gap $-20.693$~day), present when no flagged burst is excluded, is removed by \texttt{excluded\_flag}.}
\end{deluxetable*}

\subsection{Comparison with the Significance Criterion}
\label{sec:criterion-check}

No pair satisfies the $3\sigma$ or the $5\sigma$ criterion (Figure~\ref{fig:closure}).
The pair closest in time, FRB\,20190630C--IC\,190629A, is separated by $|\Delta T|=\fnFigOneDeltaTDays$~days, $\fnRatioDtThreeSigma$ times $\Delta T_{3\sigma}$, and the next closest pair, FRB\,20190701B--IC\,190629A, by $\fnNextPairDays$~days.

We also evaluate how unusual the closest pair is.
The probability that the null hypothesis produces a pair at least as close, $p(\mathrm{TS_{obs}})$, already includes the multiplicity of the $\fnNAlerts$ alerts, so no further trials correction is applied.
Since this separation lies beyond the closure breakdown (Section~\ref{sec:blind-sens}), we use the Monte Carlo probability, $p=\fnPMCObs$ ($\fnSigmaMCObs\,\sigma$), and the closest pair is consistent with a chance coincidence.
Both of the two closest pairs involve IC\,190629A, the alert with the largest share of the expected number of chance coincidences (Section~\ref{sec:blind-sens}; Figure~\ref{fig:qi_DEC}).
How these results depend on the sample definition is examined in Section~\ref{sec:robust}.

\section{Population Stacking Limits}
\label{sec:limits}
\label{sec:stack-design}

We derive 90\% upper limits on the neutrino-to-radio luminosity ratio $\xi$ of the FRB population by stacking the $\fnNFRBLimitsXi$ non-repeating bursts that have a catalogue fluence, and on the neutrino fluence per burst, for comparison with previous works, by stacking all $\fnNFRBNonRepeating$ non-repeating bursts with equal fluence (Table~\ref{tab:frbsample}).
The stacking analysis is configured from catalogue properties alone, independently of the observed pairs.
An FRB and a neutrino alert are coincident if the FRB lies inside the 90\% localization region of the alert and $|t_j-t_i|\le\Delta t_{{\rm FRB},j}$, where $\Delta t_{{\rm FRB},j}$ is the burst duration given in the catalogue; we call this interval the burst-duration window.
Burst durations are of order milliseconds, so the expected number of chance coincidences is negligible: it is $\langle N\rangle=\fnNMeanAtThreeSigma$ even for $\Delta T_{3\sigma}$, and it is proportional to $\Delta T$ (Equation~(\ref{eq:N})).
For zero observed events and zero background, the one-sided 90\% upper limit on the expected count is therefore $\mu_{90}=-\ln0.1\simeq\fnMuNinety$.

No pair lies within $\Delta T_{3\sigma}$ (Section~\ref{sec:results}), and the burst-duration window is much shorter, so $N_{\rm obs}=0$.

\subsection{Neutrino-to-Radio Luminosity Ratio}
\label{sec:xi}

To investigate the photo-hadronic neutrino emission predicted for FRBs \citep{Metzger_2020,Qu_2022,Shimoda_2024}, we constrain the neutrino-to-radio luminosity ratio $\xi$.
We define it in the source frame as
\begin{equation}
  \xi \equiv \frac{L_\nu}{L_{\rm FRB}} = \frac{E_\nu^{\rm iso}}{E_{\rm FRB}^{\rm iso}}\,\frac{\Delta t_{\rm FRB}}{\Delta t_\nu},
  \label{eq:xi}
\end{equation}
where $L=E^{\rm iso}/\Delta t$ is the isotropic-equivalent luminosity, integrated over the observed energy range 1~TeV--1~PeV for neutrinos and over the observed bandwidth of the FRB for the radio emission.
We assume $\Delta t_\nu=\Delta t_{\rm FRB}$, the catalogue burst duration, so that $\xi$ is the ratio of the neutrino and radio energies \citep[cf.][]{Metzger_2020}.
We take $\xi$ to be the mean over the burst population, weighted by $w_i$ (Equation~(\ref{eq:N-xi})), which does not imply the same intrinsic ratio for every burst.
\citet{Yamasaki_2016} uses the same symbol for the ratio of monochromatic luminosities of GeV gamma-ray counterparts, so their values are not directly comparable to our band-integrated ones.

For both neutrinos and radio emission, the source-frame isotropic-equivalent energy is $E^{\rm iso}=4\pi D_L^2(z)\,\mathcal{F}^{\rm obs}/(1+z)$, where $\mathcal{F}^{\rm obs}$ is the observed energy fluence and $D_L$ is the luminosity distance.
We obtain the radio fluence $\mathcal{F}^{\rm obs}_{{\rm FRB},i}$ by multiplying the catalogue fluence density $\mathcal{F}^{\rm obs}_{\nu,{\rm FRB},i}$ (Jy~ms) by the effective bandwidth $\Delta\nu$.
Because the factor $4\pi D_L^2/(1+z)$ is common to both, the observed neutrino energy fluence is
\begin{equation}
  \Phi^{\rm obs}_{\nu,i} = \xi\,\mathcal{F}^{\rm obs}_{{\rm FRB},i},
  \label{eq:fnu}
\end{equation}
and neither the redshift nor its estimate from the DM enters the limits, apart from the K-correction discussed in Section~\ref{sec:kcorr}.

\subsection{Expected Neutrino Counts}
\label{sec:counts}

The expected number of neutrinos detected from an $i$-th FRB at declination $\delta_i$ is
\begin{equation}
  \lambda_i(\xi)=\int_{1~{\rm TeV}}^{1~{\rm PeV}}\frac{A_{\rm eff}(\epsilon^{\rm obs},\delta_i)}{\epsilon^{\rm obs}}\,\frac{d\Phi^{\rm obs}_{\nu,i}}{d\epsilon^{\rm obs}}\,d\epsilon^{\rm obs},
  \label{eq:lambda}
\end{equation}
where $\epsilon^{\rm obs}$ is the observed neutrino energy, $d\Phi^{\rm obs}_{\nu,i}/d\epsilon^{\rm obs}$ is the energy fluence per unit energy, whose integral over 1~TeV--1~PeV is $\Phi^{\rm obs}_{\nu,i}$, and $A_{\rm eff}$ is the IceCube muon-neutrino effective area for the Gold and Bronze selections of ICECAT1 \citep{Abbasi_2023b}.
The integration range follows previous works \citep{Fahey_2017,Albert_2018a,Aartsen_2018c,Aartsen_2020a,Abbasi_2023c}.
To account for the uncertain neutrino emission mechanism, we evaluate power-law number spectra $dN/d\epsilon^{\rm obs}\propto(\epsilon^{\rm obs})^{-\gamma}$ with $\gamma=1.0$--3.0 in steps of 0.5, so that $\gamma=2$ gives equal energy per logarithmic energy interval.
This range brackets efficient particle acceleration such as diffusive shock acceleration ($\gamma\sim2$) and softer spectra from radiative cooling, particle escape or transport effects \citep[e.g.][]{Becker_2008,Kurahashi_2022}.
From Equations~(\ref{eq:fnu}) and (\ref{eq:lambda}), the total number of neutrinos expected from the $\fnNFRBLimitsXi$ bursts is
\begin{equation}
  N(\xi)=\sum_i\lambda_i=\xi\sum_i w_i, \qquad w_i=\mathcal{F}^{\rm obs}_{{\rm FRB},i}K_i,
  \label{eq:N-xi}
\end{equation}
where $K_i=\lambda_i/\Phi^{\rm obs}_{\nu,i}$ is the expected number of neutrinos per unit fluence.
With $N_{\rm obs}=0$ and negligible background, the 90\% limit follows from $N(\xi_{90})=\mu_{90}$.

\subsection{Upper Limits}
\label{sec:xi-limits}

Table~\ref{tab:xi} lists the 90\% upper limits on $\xi$ and on the neutrino fluence per burst, $\Phi^{\rm obs}_{\nu,90}$, for each $\gamma$; the latter assumes equal fluence for all bursts (Table~\ref{tab:prior} in Section~\ref{sec:prior}).
The limit on $\xi$ weakens by more than an order of magnitude from $\gamma=1$ to 3, because softer spectra place more of the energy fluence at energies where the effective area is small.

\begin{deluxetable}{cccc}
\tabletypesize{\small}
\tablewidth{0pt}
\tablecaption{90\% Upper Limits on $\xi$ and on the Neutrino Fluence per Burst\label{tab:xi}}
\tablehead{
\colhead{$\gamma$} & \colhead{$\xi_{90}$} & \colhead{$\Phi_{\rm pivot}$ (GeV cm$^{-2}$)} & \colhead{$E^{\mathrm{iso}}_{\nu,90}$ (erg)}
}
\startdata
1.0 & $1.4\times10^{11}$ & $1.77\times10^{-4}$ & $\sim10^{49\mbox{--}52}$\\
1.5 & $1.6\times10^{11}$ & $3.29\times10^{-4}$ & $\sim10^{49\mbox{--}52}$\\
2.0 & $3.0\times10^{11}$ & $5.28\times10^{-4}$ & $\sim10^{49\mbox{--}52}$\\
2.5 & $1.2\times10^{12}$ & $7.12\times10^{-4}$ & $\sim10^{50\mbox{--}53}$\\
3.0 & $6.8\times10^{12}$ & $8.01\times10^{-4}$ & $\sim10^{50\mbox{--}53}$\\
\enddata
\tablecomments{Stacked over the $\fnNFRBLimitsXi$ bursts of the baseline sample, for neutrino number spectra $dN/d\epsilon^{\rm obs}\propto(\epsilon^{\rm obs})^{-\gamma}$ over 1~TeV--1~PeV. $\Phi_{\rm pivot}$ is $(\epsilon_0^{\rm obs})^2dN/d\epsilon^{\rm obs}$ at $\epsilon_0^{\rm obs}=100$~TeV, for equal fluence for all $\fnNFRBValidCoordinates$ bursts with a valid declination. $E^{\rm iso}_{\nu,90}$ is $\xi_{90}$ times a typical radio energy of $10^{38}$--$10^{41}$~erg.}
\end{deluxetable}

The weights $w_i$ are dominated by bursts with large fluences: for $\gamma=2$, the effective number of bursts, $N_{\rm eff}=(\sum_iw_i)^2/\sum_iw_i^2$, is $\fnNeffW$ for the $\fnNFRBLimitsXi$ bursts, compared with $\fnNeffK$ for the $\fnNFRBValidCoordinates$ bursts of the per-burst limit, which depends on $K_i$ alone.
The burst with the largest weight, FRB\,20200320A (fluence 123~Jy~ms, $\delta=16^\circ$), carries 3.1\% of $\sum_iw_i$ for $\gamma=2$, so the stacked limit on $\xi$ is only a factor of 33 tighter than the limit from this burst alone (30--36 for $\gamma=1$--3).
The limit on $\xi$ therefore reflects mainly a small fraction of the sample, whereas the per-burst limit reflects most of it.

\subsection{Systematic Uncertainties}
\label{sec:syst}

We consider two effects on the limit on $\xi$: the K-correction (Section~\ref{sec:kcorr}) and the fluence calibration (Section~\ref{sec:fluence-cal}).

\subsubsection{K-Correction}
\label{sec:kcorr}

Equation~(\ref{eq:fnu}) is exact for the source-frame bands that correspond to the observed ones.
For fixed source-frame bands, the K-correction would change $\xi$ by $(1+z)^{\gamma-\alpha-1}$, where $\mathcal{F}_\nu\propto\nu^{-\alpha}$ and $\gamma$ is the neutrino spectral index.
For $\fnRedshiftFractionBelowReferencePct$\% of our bursts, which lie at $z<\fnRedshiftReference$ in redshifts estimated from the extragalactic DM \citep[following][]{Hashimoto_2022}, this is at most a factor of $\fnKCorrMaxFactorAtReference$ in either direction ($\alpha=0$--2, $\gamma=1$--3), and we neglect it because we interpret $\xi_{90}$ only to order of magnitude.

\subsubsection{Fluence Calibration}
\label{sec:fluence-cal}

We examine how the fluence calibration affects the limit on $\xi$, which depends on the burst fluences through the weights $w_i\propto\mathcal{F}^{\rm obs}_{{\rm FRB},i}$.
CHIME/FRB catalogue fluences are derived from intensity data and are lower limits, because the position of a burst within the formed beams is too uncertain to correct for the beam response \citep{Hashimoto_2022,Shin_2023,Tang_2023}, so the limits of Table~\ref{tab:xi} are in general conservative.

We estimate the effect with the $\fnNBasebandUsed$ bursts of our sample that are in the first CHIME/FRB baseband catalogue \citep{Amiri_2024a}, whose fluences are measured at the baseband positions.
The baseband catalogue raises the fluences of about $\fnBasebandFracPct$\% of the sample by a median factor of $\fnCRatioMedian$, and lowers them for only $\fnNCRatioBelowOne$ bursts.
At $\gamma=2$, the summed weight of these bursts increases by a factor of $\fnGPhot$, but they carry only a fraction $\fnSSamp$ of the total weight, so the limit would improve by less than a factor of 2 if only these bursts were corrected.

We do not extrapolate this to the full sample and retain the intensity-based limits, because the gain rests on few bursts (the effective number of bursts is $\fnNeffWRowC$) and the subsample need not be representative.
The fluence calibration therefore does not change the order of magnitude of the limits for the bursts we can test.


\section{Discussion}
\label{sec:Discussion}

\subsection{Dependence on the Sample Definition}
\label{sec:robust}

Table~\ref{table:robust_flags} compares the baseline sample of Section~\ref{sec:data}, used in Sections~\ref{sec:results} and~\ref{sec:limits}, with the sample without flag exclusion and with the sample that includes the repeaters.
The closest pair is FRB\,20190630C--IC\,190629A, separated by $\fnFigOneDeltaTDays$~days, in all three samples.
The sample without flag exclusion has one more pair within $\pm\fnPairWindowDays$~days, FRB\,20190609A--IC\,190629A, which \texttt{excluded\_flag} removes from the baseline.
The $3\sigma$ and $5\sigma$ windows recomputed for each sample differ from the baseline by less than the systematic uncertainty from the rate-bin width (Section~\ref{sec:blind-sens}), and $p$ changes only slightly.
No pair in any sample satisfies the significance criterion, so the absence of a coincidence does not depend on how the sample is defined.
The stacking limit for $\gamma=2$ is less than 10\% stronger when the flagged bursts are included, because the sample then contains more bursts.

\begin{deluxetable}{lccc}
\tabletypesize{\small}
\tablewidth{0pt}
\tablecaption{Dependence on the FRB Sample}
\label{table:robust_flags}
\tablehead{
\colhead{} & \colhead{Baseline} & \colhead{Flags incl.} & \colhead{Repeaters incl.}
}
\startdata
Pairs ($<30$~d) & 11 & 12 & 11\\
$\Delta T_{3\sigma}$ (s) & 350.1 & 341.9 & 341.0\\
$\Delta T_{5\sigma}$ (ms) & 74.30 & 72.55 & 72.36\\
$p$ (analytic) & 0.268 & 0.273 & 0.274\\
$p$ (MC) & 0.254 & 0.263 & ---\\
$\xi_{90}$ ($10^{11}$) & 3.02 & 2.74 & ---\\
$\Phi_{\rm band}$ ($10^{-3}$) & 3.65 & 3.35 & ---\\
$\Phi_{\rm pivot}$ ($10^{-4}$) & 5.28 & 4.85 & ---\\
\enddata
\tablecomments{The baseline excludes quality-flagged bursts and repeaters; Flags incl. keeps the flagged bursts. Dashes: not computed. $\xi_{90}$, $\Phi_{\rm band}$ (energy fluence over 1~TeV--1~PeV) and $\Phi_{\rm pivot}$ are the stacking limits for $\gamma=2$ (Table~\ref{tab:xi}); $\Phi$ in GeV~cm$^{-2}$.}
\end{deluxetable}

\subsection{Comparison with Previous Works}
\label{sec:prior}

Table~\ref{tab:prior} compares our limit on the neutrino fluence per burst with those of previous works, for a neutrino number spectrum $dN/d\epsilon^{\rm obs}\propto(\epsilon^{\rm obs})^{-2}$ at the pivot energy of $\fnPivotEnergyTeV$~TeV, the convention of \citet{Aartsen_2020a}.
Assuming equal fluence for all bursts, our limit is $\fnFluencePivotTwo$~GeV~cm$^{-2}$, the lowest in the table and a factor of about 4 below the stacking limits of \citet{Aartsen_2018c} and \citet{Aartsen_2020a}.
The larger sample outweighs the lower per-burst sensitivity of the ICECAT1 alerts.
Because the studies differ in hypothesis, weighting, hemisphere, event selection and time window, the comparison is indicative.

The diffuse neutrino flux measured by IceCube gives an indirect limit on the fluence per burst if it is shared among the FRB population \citep{Fahey_2017,Abbasi_2023c}.
For an all-sky rate of 820 bursts per day and a spectrum $(\epsilon^{\rm obs})^{-2.53}$, this limit is about a factor of 3 below ours \citep{Abbasi_2023c}.
However, it bounds the total neutrino emission of the population regardless of the arrival times, and cannot test whether the neutrinos coincide in time with FRBs, as our search does.

\begin{deluxetable*}{lccccc}
\tablewidth{0pt}
\tablecaption{Comparison of Upper Limits on the Neutrino Fluence per Burst\label{tab:prior}}
\tablehead{
\colhead{Study} & \colhead{Neutrino data} & \colhead{$N_{\rm FRB}$} & \colhead{Time window} & \colhead{Limit (GeV\,cm$^{-2}$)} & \colhead{Ratio}
}
\startdata
\citet{Fahey_2017} & IceCube, 1\,yr track & 4 & 1\,day & $2.66\times10^{-2}$\tablenotemark{a} & \fnPriorRatioFahey \\
\citet{Albert_2018a} & ANTARES & 12 & $T_0\pm6$\,h & 1.1 & \fnPriorRatioAlbert \\
\citet{Aartsen_2018c} & IceCube, 6\,yr track, northern stacking & 29 & 10\,ms & $2.1\times10^{-3}$ & \fnPriorRatioAartsenA \\
\citet{Aartsen_2020a} & IceCube, track, stacking & 28 & $<10$\,s & $2.0\times10^{-3}$ & \fnPriorRatioAartsenB \\
\citet{Abbasi_2023c} & IceCube, 7\,yr cascade & 22 & one-sided\tablenotemark{b} & $2.0\times10^{-2}$ & \fnPriorRatioAbbasi \\
This work & ICECAT1 alerts & $\fnNFRBValidCoordinates$ & $\Delta t_{\rm FRB}$ & $\fnFluencePivotTwo$ & 1 \\
\enddata
\tablecomments{Limits at 90\% confidence for a spectrum $dN/d\epsilon^{\rm obs}\propto(\epsilon^{\rm obs})^{-2}$, expressed as $(\epsilon_0^{\rm obs})^2dN/d\epsilon^{\rm obs}$ at $\epsilon_0^{\rm obs}=100$~TeV. Ratio: the limit divided by that of this work.}
\tablenotetext{a}{Converted from the energy fluence over 1~TeV--1~PeV by dividing by $\ln10^3$.}
\tablenotetext{b}{One-sided search in $[t_{\rm FRB},t_{\rm FRB}+\Delta T]$, not directly comparable to the two-sided window of this work.}
\end{deluxetable*}

\subsection{Implications for Neutrino Emission Models}
\label{subsec:models}

Magnetar-driven shock models predict neutrino energies of $10^{35}$--$10^{44}$~erg, corresponding to $\xi\sim10^{-4}$--$10^{-1}$ \citep{Metzger_2020}.
Our limits for $\gamma=2$, $\xi\lesssim3\times10^{11}$ and $\sim10^{49}$--$10^{52}$~erg (Table~\ref{tab:xi}), exceed these predictions by more than ten and five orders of magnitude, respectively; the energy gap is smaller because the upper end of the prediction assumes the most energetic FRBs.
Moreover, only neutrinos arriving within the burst duration are counted (Section~\ref{sec:xi}), so the limits would be weaker still for emission delayed by up to $\sim10^{3}$~s.
The present data therefore do not constrain these models and are relevant only for scenarios with much more energetic hadronic outflows.

The Galactic FRB\,20200428D from the magnetar SGR\,1935+2154 gives a much tighter limit because of its proximity: the IceCube fluence limit of $5\times10^{-2}$~GeV~cm$^{-2}$ quoted by \citet{Metzger_2020} corresponds to $\sim10^{42}$~erg at $\sim10$~kpc, i.e.\ $\xi\lesssim10^{7}$--$10^{8}$ for a radio energy of $10^{34}$--$10^{35}$~erg.
This is a single flare, with a radio energy several orders of magnitude below typical cosmological FRBs, and it remains more than seven orders of magnitude above the predicted $\xi$.

\subsection{Future Prospects}
\label{sec:prospects}

The stacked limit scales as $\xi_{90}\propto(\sum_i F_{{\rm FRB},i}K_i)^{-1}$, and a few bright bursts dominate the weights (Section~\ref{sec:xi-limits}).
Future gains therefore come mainly from additional bright bursts at declinations where IceCube is sensitive.
Extending the FRB sample to the full second CHIME/FRB catalogue (to 2023 September) would increase the number of bursts by at most a factor of $\approx 2.5$ (4539 versus 1823), provided that neutrino data covering the extended period are available. The event-level IceTracks-DR2 \citep{Abbasi_2026}, which extends to 2022 May and provides effective areas, could in principle be used for this purpose; we leave such a search to future work.
The Bustling Universe Radio Survey Telescope in Taiwan (BURSTT), in its 256-antenna configuration, is expected to detect $\sim$100 FRBs per year above 100~Jy~ms \citep{Lin_2022}, comparable to the brightest burst in our sample (123~Jy~ms).

Even a gain of two orders of magnitude, however, falls far short of what the models require.
Stacking our 1423 bursts at 1~Gpc with equal fluences constrains the mean isotropic neutrino energy only to $E_\nu\approx6\times10^{50}$~erg for $\gamma=2$.
Reaching $10^{44}$~erg, even the most optimistic prediction, would require a sample of order $10^{7}$ times larger ($\sim10^{10}$ bursts), and $\sim10^{8}$ bursts even at 100~Mpc.

A nearby burst lowers the limit on $\xi$ far faster than a larger sample: FRB\,20200428D alone is three to four orders of magnitude tighter than our stack, yet still far above the predicted values (Section~\ref{subsec:models}).
Reaching them requires an exceptional event; \citet{Metzger_2020} identify a giant flare from a nearby Galactic magnetar as the most promising case.



\section{Summary}
\label{sec:summary}

We searched for temporal and spatial coincidences between the $\fnNFRBNonRepeating$ non-repeating FRBs of the Second CHIME/FRB catalogue and the $\fnNAlerts$ alerts of ICECAT1, and find no statistically significant association: no pair lies within the $3\sigma$ and $5\sigma$ windows $\Delta T_{3\sigma}=\fnDtThreeSigma$~s and $\Delta T_{5\sigma}=\fnDtFiveSigma$~s.
The closest pair, FRB\,20190630C--IC\,190629A, is $\fnFigOneDeltaTDays$~days apart and consistent with a chance coincidence ($p=\fnPMCObs$, $\fnSigmaMCObs\,\sigma$), independently of the sample definition (Section~\ref{sec:robust}).

Stacking the $\fnNFRBLimitsXi$ bursts, we obtain 90\% upper limits on the neutrino-to-radio luminosity ratio of $\xi\lesssim\fnXiULOne$ for $\gamma=1.0$ to $\xi\lesssim\fnXiULThree$ for $\gamma=3.0$, assuming neutrino emission within the burst duration (Section~\ref{sec:limits}).
For a neutrino number spectrum $dN/d\epsilon^{\rm obs}\propto(\epsilon^{\rm obs})^{-2}$, the limit on the neutrino fluence per burst is about a factor of 4 below those of previous stacking searches (Section~\ref{sec:prior}).
These limits lie more than five orders of magnitude above the predictions of magnetar-driven shock models, so the present data do not test them (Section~\ref{subsec:models}).
Larger cosmological samples are unlikely to close this gap; reaching the predictions requires exceptional nearby events, such as a giant flare from a Galactic magnetar, for which the blind criterion and the limits presented here provide the baseline (Section~\ref{sec:prospects}).

\begin{acknowledgments}
This study was supported by JST SPRING (JPMJSP2108), and International Graduate Program for Excellence in Earth-Space Science (IGPEES), a World-leading Innovative Graduate Study (WINGS) Program, the University of Tokyo, and International Internship Pilot Program (IIPP), provided by the National Science and Technology Council (NSTC) in Taiwan. We also gratefully acknowledge the CHIME/FRB Collaboration and the IceCube Collaboration for providing the high-quality catalogs essential to this research. Their efforts and open data sharing have been invaluable to the advancement of our work.
TH acknowledges the support of the National Science and Technology Council of Taiwan (NSTC) through grants 113-2112-M-005-009-MY3, 113-2123-M-001-008-, and 111-2112-M-005-018-MY3 and the Ministry of Education of Taiwan through the grant 113RD109. SY acknowledges the support from the NSTC through grant numbers 113-2112-M-005-007-MY3 and 113-2811-M-005-006-. 
This work is supported by the Grants-in-Aid for Scientific Research Nos. 25KJ0024, 25K17378 (T.W.) from the Ministry of Education, Culture, Sports, Science and Technology (MEXT) of Japan.
\end{acknowledgments}

\bibliographystyle{aasjournalv7}
\bibliography{apj}

\label{lastpage}
\end{document}